\RequirePackage{fix-cm}
\documentclass[smallextended]{svjour3}       
\smartqed  
\usepackage{graphicx}
\usepackage{epstopdf}
\usepackage{amsmath}
\usepackage{multirow}
\usepackage{array}
\begin{document}

\title{Interaction of antiproton with nuclei \thanks{
J. M. was supported by GACR Grant No. P203/12/2126. J. H. acknowledges financial support from CTU-SGS Grant No. SGS13/216/OHK4/3T/14. 
}
}


\author{J. Hrt\'{a}nkov\'{a} \and
J. Mare\v{s}
}


\institute{J. Hrt\'{a}nkov\'{a}  \at
              Nuclear Physics Institute, 25068 \v{R}e\v{z}, Czech Republic  \\
\emph{and} Czech Technical University in Prague, Faculty of Nuclear Sciences and Physical 
Engineering, B\v{r}ehov\'{a} 7, 115 19 Prague 1, Czech Republic\\ 
              Tel.: +420 266 173 283\\
              Fax: +420 220 940 165\\
              \email{hrtankova@ujf.cas.cz}           \\
                    \and J. Mare\v{s}
 \at 
Nuclear Physics Institute, 25068 \v{R}e\v{z}, Czech Republic 
}

\date{Received: date / Accepted: date}

\maketitle

\begin{abstract}
We performed fully self-consistent calculations of $\bar{p}$--nuclear bound states within the relativistic mean-field (RMF) model. The G-parity motivated $\bar{p}$--meson coupling constants were adjusted to yield potentials consistent with $\bar{p}$--atom data. We confirmed large polarization effects of the nuclear core caused by the presence of the antiproton. The $\bar{p}$ absorption in the nucleus was incorporated by means of the imaginary part of a phenomenological optical potential. The phase space reduction for the $\bar{p}$ annihilation products was taken into account. The corresponding $\bar{p}$ width in the medium significantly decreases, however, it still remains considerable for the $\bar{p}$ potential consistent with experimental data.

\end{abstract}

\section{Introduction}
\label{intro}
The study of the antiproton--nucleus interaction is an interesting issue which has attracted renewed interest in recent years at the prospect of future experiments at the FAIR facility. The $\bar{p}$--nuclear bound states and the possibility of their formation have been studied in refs.~\cite{Mishustin,lari mish satarov}. These considerations are supported by a strongly attractive potential that the $\bar{p}$ feels in the nuclear medium. Within the RMF approach the real part of the $\bar{p}$--nucleus potential derived using the G-parity transformation is Re$V_{\bar{p}} \simeq-650$~MeV deep at normal nuclear density. However, the experiments with $\bar{p}$ atoms \cite{mares} and $\bar{p}$ scattering off nuclei at low energies \cite{antiNN interaction} favor shallower real part of the $\bar{p}$--nucleus potential in the range of $-(100 - 300)$~MeV in the nuclear interior. An important aspect of the $\bar{p}$--nucleus interaction is $\bar{p}$ annihilation which appears to be the dominant part of the interaction. Nevertheless, the phase space for the annihilation products should be significantly suppressed for the antiproton bound deeply in the nuclear medium, which could lead to the relatively long living $\bar{p}$ inside the nucleus~\cite{Mishustin}.

In this contribution, we report on our recent fully self-consistent calculations of $\bar{p}$--nuclear bound states including $\bar{p}$ absorption in a nucleus. The calculations are performed within the RMF model \cite{Walecka}. Dynamical effects in the nuclear core caused by the antiproton and the phase space suppression for the $\bar{p}$ annihilation products are studied for various nuclei.

In Section 2, we briefly introduce the underlying model. Few selected representative results of our calculations are discussed in Section 3.

\section{Model}
\label{sec:1}
The $\bar{p}$--nucleus interaction is described within the RMF approach. The interaction among (anti)nucleons is mediated by the exchange of the scalar ($\sigma$)
 and vector ($\omega_{\mu}$, $\vec{\rho}_\mu$) meson fields, and the massless photon field $A_{\mu}$. The standard Lagrangian density $\mathcal{L}_N$ for nucleonic 
sector is extended by the Lagrangian density $\mathcal{L}_{\bar{p}}$ describing the antiproton 
interaction with the nuclear medium (see ref.~\cite{jarka} for details). The variational principle yields the equations of motion 
for the hadron fields involved. The Dirac equations for nucleons and antiproton read:  
\begin{equation} \label{Dirac antiproton}
[-i\vec{\alpha}\vec{\nabla} +\beta(m_j + S_j) + V_j]\psi_j^{\alpha}=\epsilon_j^{\alpha} \psi_j^{\alpha}, 
\quad j=N,\bar{p}~,
\end{equation}
where
\begin{equation}
S_j=g_{\sigma j}\sigma, \quad V_j=g_{\omega j} \omega_0 + g_{\rho j}\rho_0 \tau_3 + e_j \frac{1+\tau_3}{2}A_0
\end{equation}
are the scalar and vector potentials. Here, $m_j$ stands for (anti)nucleon mass; $g_{\sigma j}, g_{\omega j}, g_{\rho j}$, and $e_j$ are the (anti)nucleon couplings to corresponding fields, and $\alpha$ denotes single particle states. The Klein--Gordon equations for the boson fields acquire additional source terms due to the presence of $\bar{p}$:
\begin{equation}
\begin{split} \label{meson eq}
(-\triangle + m_\sigma^2+ g_2\sigma + g_3\sigma^2)\sigma&=- g_{\sigma N} \rho_{SN}-g_{\sigma \bar{p}} 
\rho_{S \bar{p}}~, \\
(-\triangle + m_\omega^2 +d\omega^2_0)\omega_0&= g_{\omega N}\rho_{VN} +g_{\omega \bar{p}} \rho_{V\bar{p}}~, \\
(-\triangle + m_\rho^2)\rho_0&= g_{\rho N}\rho_{IN} +g_{\rho \bar{p}}\rho_{I \bar{p}}~, \\
-\triangle A_0&= e_N \rho_{QN}+e_{\bar{p}}\rho_{Q\bar{p}}~,
\end{split}
\end{equation}
where $\rho_{\text{S}j}, \rho_{\text{V}j}, \rho_{\text{I}j}$ and $\rho_{\text{Q}j}$  are the scalar, 
vector, isovector, and charge densities, respectively, and $m_\sigma, m_\omega, m_\rho$ are the masses of the
considered mesons. 
In this work, the nucleon--meson coupling constants and meson masses were adopted from the nonlinear RMF model TM1(2) \cite{Toki} for heavy (light) nuclei. The system of the coupled Dirac  \eqref{Dirac antiproton} and Klein--Gordon \eqref{meson eq} equations is solved fully self-consistently by iterative procedure.
\\

In the RMF model, the nucleon in a nucleus moves in mean fields created by all nucleons, i.\ e., the nucleon feels repulsion as well as attraction also from itself. In ordinary nuclei this nucleon self-interaction has only a minor ($1/\text{A}$) effect. However, the potential acting on $\bar{p}$ in a nucleus is much deeper and the impact of the $\bar{p}$ self-interaction could become pronounced. In order to exclude this unphysical $\bar{p}$ self-interaction we omitted the antiproton source terms in the Klein--Gordon equations for the boson fields acting on the $\bar{p}$.\footnote{It is to be noted that the self-interaction is directly subtracted in the Hartree--Fock formalism.}

The $\bar{p}$--nucleus interaction is constructed from the $p$--nucleus interaction with the help of the 
G-parity transformation: the vector potential generated by the $\omega$ meson exchange thus changes its 
sign and becomes attractive. As a consequence, the total $\bar{p}$ potential will be strongly 
attractive. However, the G-parity transformation should be regarded as a mere starting point to determine the $\bar{p}$--meson 
coupling constants. Various many-body effects, as well as the presence of strong annihilation channels 
could cause significant deviations from the G-parity values in the nuclear medium. Therefore, we introduce a scaling factor $\xi \in \langle0,1\rangle$ for the $\bar{p}$--meson coupling constants \cite{Mishustin}:
\begin{equation}
 g_{\sigma \bar{p}}=\xi\, g_{\sigma N}, \quad g_{\omega \bar{p}}=-\xi\, g_{\omega N}, \quad g_{\rho \bar{p}}=\xi\, g_{\rho N}~.
\end{equation}

\bigskip
The $\bar{p}$ annihilation in the nuclear medium is described by the imaginary part of the optical potential in a `$t\rho$' 
form adopted from optical model phenomenology \cite{mares}:
\begin{equation}
2\mu {\rm Im}V_{\text{opt}}(r)=-4 \pi \left(1+ \frac{\mu}{m_N}\frac{A-1}{A} 
\right){\rm Im}b_0 \rho(r)~,
\end{equation}
where $\mu$ is the $\bar{p}$--nucleus reduced mass. While the density $\rho(r)$ was treated as a dynamical 
quantity evaluated within the RMF model, the parameter Im$b_0=1.9$~fm was determined by fitting the $\bar{p}$ atom 
data \cite{mares}. 

The energy available for the $\bar{p}$ annihilation in the nuclear medium is usually expressed as 
$\sqrt{s}=~m_{\bar{p}}+m_{N}-B_{\bar{p}}-B_{N}$, where $B_{\bar{p}}$ and $B_{N}$ is the $\bar{p}$ and 
nucleon binding energy, respectively. The phase space available for the annihilation products is thus considerably suppressed for the deeply bound antiproton. 
\begin{figure}[t]
\begin{minipage}{0.48\textwidth} \vspace{-11pt}
\includegraphics[height=0.27\textheight]{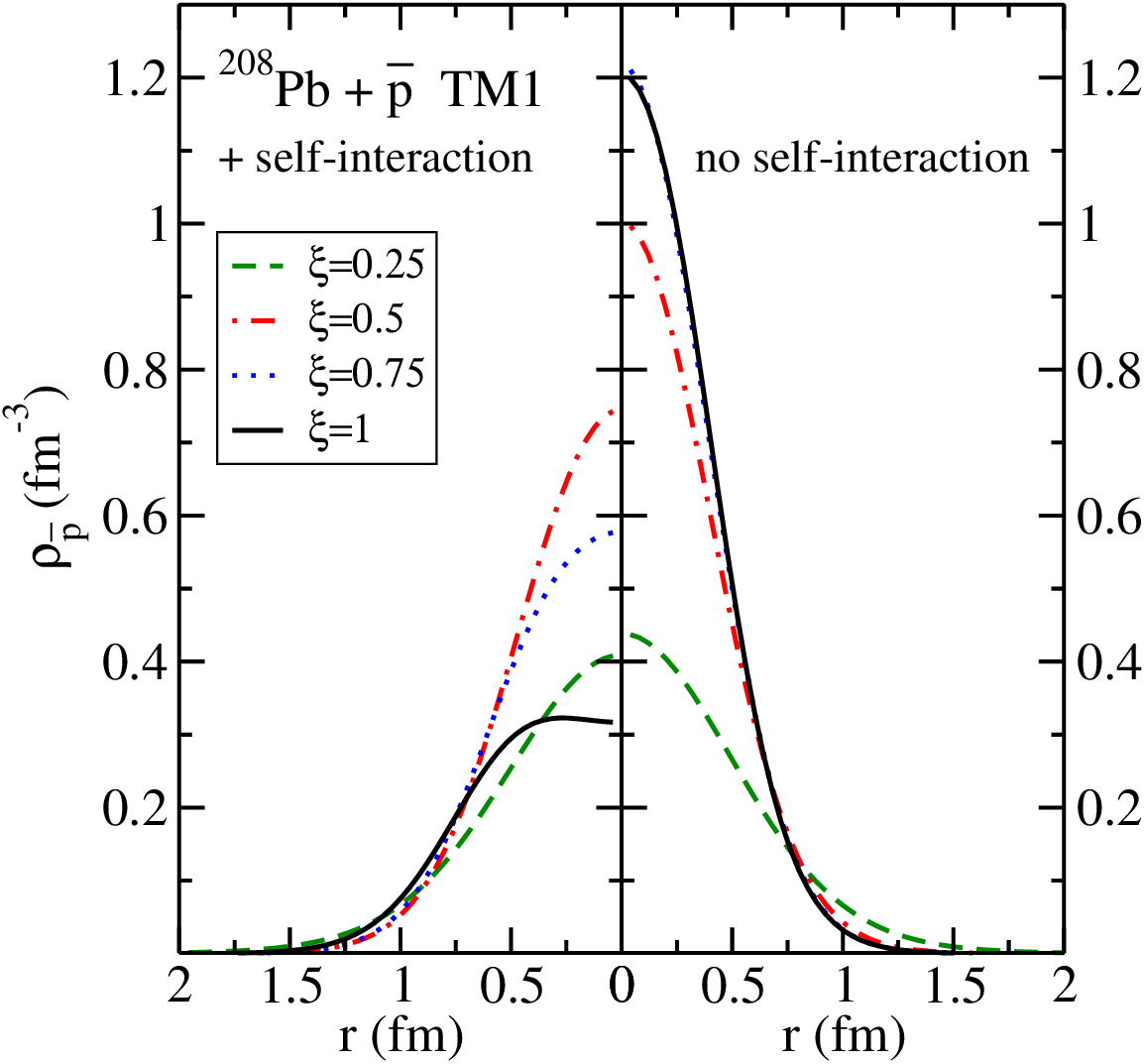}
\caption{\label{Fig.: selfinteraction}The $\bar{p}$ density distribution in $^{208}$Pb, calculated for 
different $\xi$ in the TM1 model with (left) and without (right) the $\bar{p}$ self-interaction.}
\end{minipage}\hspace{1.2pc}%
\begin{minipage}{0.48\textwidth}
\includegraphics[height=0.27\textheight]{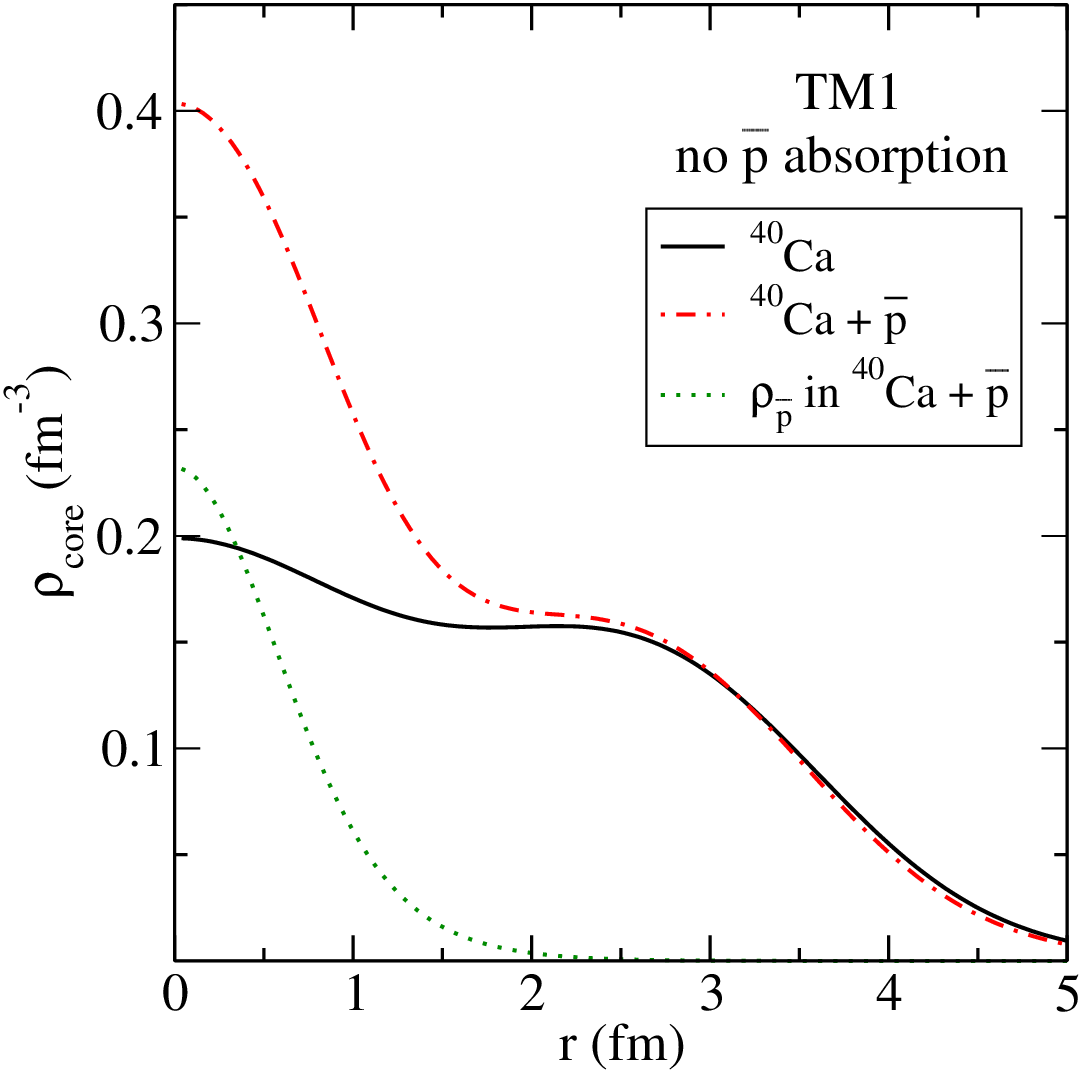}
\caption{\label{Fig.: density}The nuclear core density distribution in $^{40}$Ca and $^{40}$Ca$+{\bar{p}}$ for $\xi=0.2$, calculated in the TM1 model. The $\bar{p}$ density distribution in $^{40}$Ca$+{\bar{p}}$ (dotted line) is shown for comparison.}
\end{minipage}
\end{figure}

The phase space suppression factor $f_{\text{s}}$ for two body decay is given by \cite{pdg}
\begin{equation}
f_{\text{s}}=\frac{M^2}{s}\sqrt{\frac{[s-(m_1+m_2)^2][s-(m_1-m_2)^2]}{[M^2-(m_1+m_2)][M^2-(m_1-m_2)^2]}}\Theta(\sqrt{s}-m_1-m_2)~,
\end{equation}
where $m_1$, $m_2$ are the masses of the annihilation products and $M=m_{\bar{p}}+m_N$. For channels containing more than 2 particles in the final state the $f_{\text{s}}$ was evaluated with the help of the Monte Carlo simulation tool PLUTO~\cite{PLUTO}. 

\section{Results}
The formalism introduced above was employed in the self-consistent calculations of the $\bar{p}$ bound states in selected nuclei. First, we did not consider the $\bar{p}$ annihilation in the nuclear medium and focused on the study of the dynamical effects caused by the presence of the antiproton in the nucleus. During our calculations we noticed a pronounced effect of the $\bar{p}$ self-interaction on the calculated observables. In Fig.~\ref{Fig.: selfinteraction}, we present the $\bar{p}$ density distribution in $^{208}$Pb, calculated dynamically in the TM1 model for different values of the scaling parameter $\xi$. The central $\bar{p}$ density $\rho_{\bar{p}}(0)$ calculated including the $\bar{p}$ self-interaction (left panel) reaches its maximum for $\xi \approx 0.5$ and then starts to decrease. It is due to the interplay between the negative value of $\text{S}_{\bar{p}} - \text{V}_{\bar{p}}$ (absolute value of which increases with $\xi$), the $\bar{p}$ single particle energy, and the $\bar{p}$ rest mass, which affects the solution of the Dirac equation for the $\bar{p}$ wave function. On the other hand, when the $\bar{p}$ self-interaction is subtracted (right panel), the scalar and vector potentials are of comparable depth and the $\rho_{\bar{p}}(0)$ increases gradually with $\xi$ and saturates at much higher values. It is to be stressed that the effect of the $\bar{p}$ self-interaction is negligible for $\xi \leq 0.2$, which includes the $\bar{p}$ potentials consistent with $\bar{p}$ atom data.

Our calculations revealed large polarization of the nuclear core caused by the $\bar{p}$ in the nuclear $1s$ state. The nuclear core density in a $\bar{p}$ nucleus reaches $2-3$ times the nuclear matter density as illustrated for $^{40}$Ca in Fig. \ref{Fig.: density}. The $\bar{p}$ is localized in the center of the nucleus and the resultant nuclear core density is substantially enhanced over a small region, $r \leq 1.5$~fm.

The nucleon single particle energies are affected by the presence of the $\bar{p}$ as well. Consequently the total binding energies of $\bar{p}$ nuclei increase considerably, $B=-344.3$~MeV for $^{40}$Ca and $B=-485.4$~MeV for $^{40}$Ca+$\bar{p}$ ($\xi=0.2$).  
\begin{figure}[t]
\begin{minipage}{0.46\textwidth} \vspace{-8pt}
\includegraphics[height=0.257\textheight]{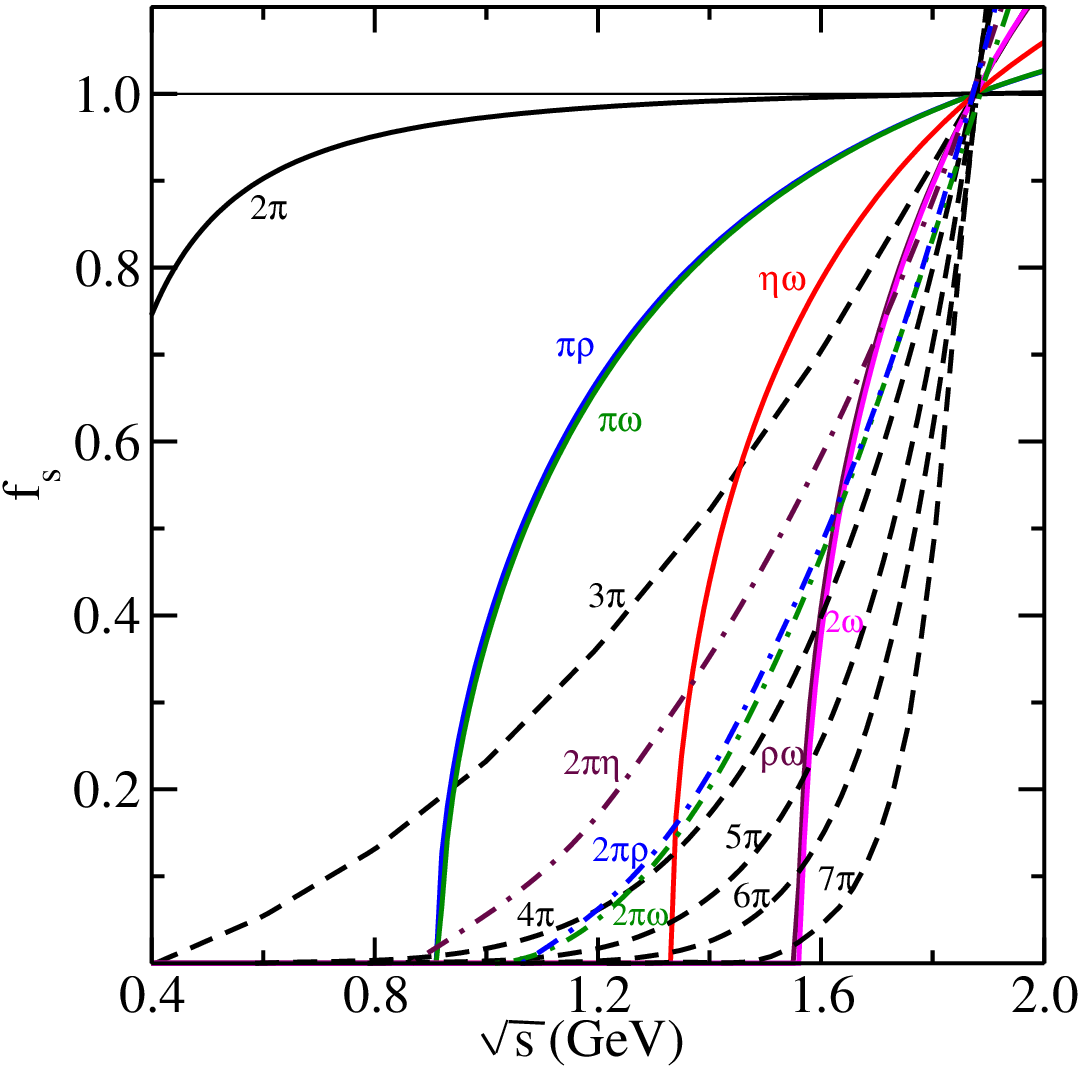}
\caption{\label{Fig.: SF}The phase space suppression factors $f_{\text{s}}$ as a function of the c.m. 
energy $\sqrt{s}$.}
\vspace*{21pt}
\end{minipage}\hspace{1.2pc}%
\begin{minipage}{0.49\textwidth}
\includegraphics[height=0.26\textheight]{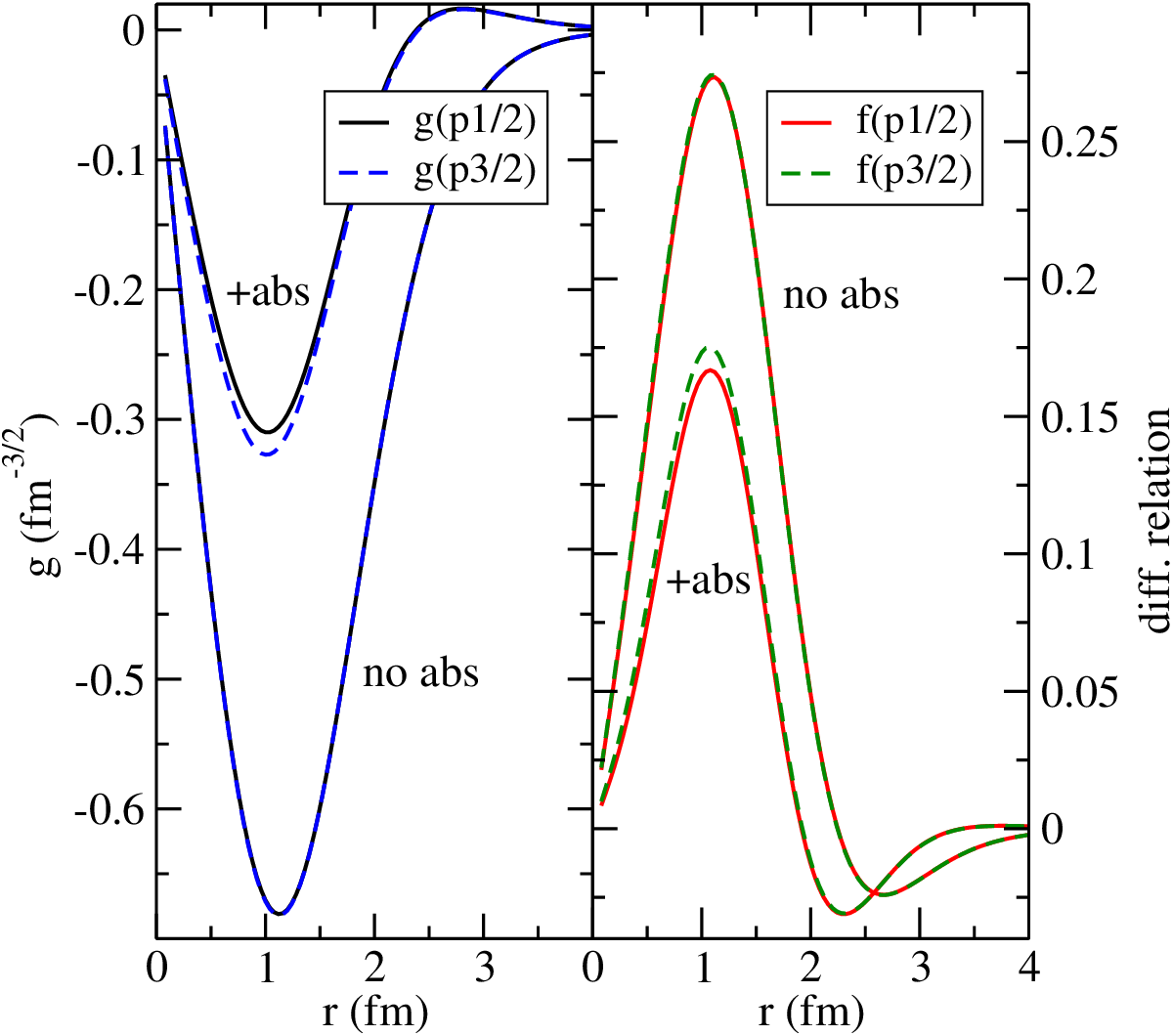}
\caption{\label{Fig.: symmetry}Upper components  $g$ (left) and the relation \eqref{diff rel} for lower components $f$ (right) of the $1 p_{1/2}$ and 
$1 p_{3/2}$ $\bar{p}$ wave functions in $^{16}$O, calculated dynamically in the TM2 model using the 
real (no abs) and complex (+ abs) $\bar{p}$ potential.} 
\end{minipage}
\end{figure}

In order to account for $\bar{p}$ annihilation, we performed calculations using the complex potential presented in Section~2. We considered the suppression of the phase space for the annihilation products. In Fig.~\ref{Fig.: SF}, the phase space suppression factors for the annihilation channels involved are presented as a function of $\sqrt{s}$. As $\sqrt{s}$ decreases due to the $\bar{p}$ and $N$ binding energies many channels become strongly suppressed or even closed. Moreover, the $\bar{p}$--nucleon annihilation takes place in the nuclear medium. Therefore, the momentum dependent term in the Mandelstam variable $s=(E_N + E_{\bar{p}})^2 - (\vec{p}_N + \vec{p}_{\bar{p}})^2$, where $E_j=m_j-B_j$, is non-negligible in contrast to the two body frame \cite{s}. Our self-consistent evaluation of $\sqrt{s}$ including $\vec{p}_{\bar{p}}$ and $\vec{p}_N$ leads to an additional downward energy shift overlooked by many previous calculations.

In Table \ref{Tab.: results}, we present the $1s$ $\bar{p}$ single particle energies and widths in $^{16}$O+$\bar{p}$, calculated using the real and complex potentials consistent with $\bar{p}$-atom data ($\xi=0.2$). To illustrate the role of the suppression factors $f_s$ we show the results of calculations without $f_s$ (`Complex'), as well as including $f_s$ for $\sqrt{s}$ due to $B_{\bar{p}}$ and $B_{N}$ (`Complex+$f_s$') and for $\sqrt{s}$ with the additional downward energy shift due to the momenta of annihilating partners (`+$\sqrt{s(\vec{p})}$'). The static calculations, which do not account for the core polarization effects, give 
approximately the same values of the $\bar{p}$ single particle energy for all cases. The single particle energies calculated 
dynamically are larger, which indicates that the polarization of the core nucleus is significant. When the phase space suppression is taken into account the $\bar{p}$ width is reduced by more than twice (compare `Complex' and `Complex+$f_s$' in the last row of Table \ref{Tab.: results}). When treating $\sqrt{s}$ self-consistently including the $\bar{p}$ and $N$ momenta, the $\bar{p}$ width is reduced by additional $\approx50$~MeV, but still remains sizeable. The corresponding lifetime of the $\bar{p}$ in the nucleus is $\simeq 1$~fm.
\begin{center}
\begin{table}[t]
\caption{\label{Tab.: results} The $1s$ single particle energies $\epsilon_{\bar{p}}$ and widths $\Gamma_{\bar{p}}$ (in MeV) in 
$^{16}$O+$\bar{p}$, calculated dynamically (Dyn) and statically (Stat) within the TM2 model using the real and complex potentials consistent with $\bar{p}$--atom data (see text for details).}
\centering
\begin{tabular}{lcccccccc}
\hline\noalign{\smallskip}
 & \multicolumn{2}{c}{Real} & \multicolumn{2}{c}{Complex} & \multicolumn{2}{l}{Complex + $f_{\text{s}}$} & \multicolumn{2}{c}{+$\sqrt{s(\vec{p})}$}  \\ \noalign{\smallskip}\hline\noalign{\smallskip}
 & Dyn & Stat & Dyn & Stat & Dyn &  Stat & Dyn & Stat  \\ \noalign{\smallskip}\hline\noalign{\smallskip}
$\epsilon_{\bar{p}}$ & ~-193.7~ & ~-137.1~ & ~-175.6~ & ~-134.6~ & ~-190.2~ & ~-136.1~ & ~-191.6~ & ~-136.3~ \\
$\Gamma_{\bar{p}}$ & ~-~ & - & ~552.3 & ~293.3 & ~232.5 & ~165.0 & ~179.9& ~144.7 \\
\noalign{\smallskip}\hline
\end{tabular}
\end{table}
\end{center} 

Finally, we discuss spin symmetry in $\bar{p}$ spectra. Static calculations with a real potential \cite{spin symmetry} revealed that antinucleon spectra in nuclei exhibit spin symmetry. We explored the $\bar{p}$ spectra considering the dynamical effects as well as the $\bar{p}$ absorption in the nucleus. In Fig. \ref{Fig.: symmetry}, the real parts of the upper components (left) and the differential relation \cite{ginocchio}
\begin{equation} \label{diff rel}
   \left(\frac{\partial}{\partial r} + \frac{\ell+2}{r}\right) f_{n_r, \ell+1/2}(r)= \left(\frac{\partial}{\partial r} - \frac{\ell-1}{r}\right) f_{n_r, \ell-1/2}(r)
\end{equation}
for the real parts of the lower components (right) of the $\bar{p}$ wave function in $1 p_{1/2}$ and $1 p_{3/2}$ states in $^{16}$O are plotted for $\xi=0.2$. We found that spin symmetry is well preserved in the $\bar{p}$ spectrum calculated fully self-consistently using the $\bar{p}$ potential consistent with the $\bar{p}$ atom data. The $\bar{p}$ annihilation causes only minor deviations. When different values of $\xi$ for the scalar and vector potentials are considered, spin symmetry holds only approximately. The deviations gradually increase with increasing difference between the scalar and vector potentials. 

\begin{acknowledgements}
We thank P. Tlust\'{y} for his assistance during Monte Carlo simulations using PLUTO, and E. Friedman, A. Gal and S. Wycech for valuable discussions. 
\end{acknowledgements}



\end{document}